\documentclass[preprint,
superscriptaddress,
amsmath,amssymb,aps,showkeys,showpacs,
twoside,final,secnumarabic,
nofootinbib]{revtex4-2}

\usepackage[paperwidth=205mm,paperheight=290mm,top=17mm,bottom=25mm,
inner=17mm,outer=17mm,
twoside]{geometry}

\usepackage{cmap} 
\usepackage[T1,T2A]{fontenc}
\usepackage[utf8]{inputenc}
\usepackage[russian,english]{babel}
\usepackage{color}
\usepackage{graphicx}
\usepackage{dcolumn}
\usepackage{bm} 
\usepackage[unicode=true,colorlinks=true,linkcolor=magenta, urlcolor=blue, citecolor = blue,breaklinks]{hyperref}
\usepackage{multirow}
\usepackage{url}
\usepackage{breakurl}
\DeclareGraphicsExtensions{.eps}

\newcount\issue
\newcount\Vol
\newcount\numb
\usepackage{fancyhdr} 
\def\Vol{\textbf{80}}
\def\numb{x}
\begin{document}
\title{ CONFERENCE SECTION \\[20pt]
Nonfactorizable charming loops in exclusive FCNC $B$ decays} 

\def\addressa{address 1}
\def\addressb{address 2}

\author{\firstname{D.I.}~\surname{Melikhov}}
\email[E-mail: ]{melikhov.dmitri@gmail.com }
\affiliation{D.~V.~Skobeltsyn Institute of Nuclear Physics, M.~V.~Lomonosov Moscow State University, 119991, Moscow, Russia}
\affiliation{Joint Institute for Nuclear Research, 141980 Dubna, Russia}
\received{xx.xx.2025}
\revised{xx.xx.2025}
\accepted{xx.xx.2025}

\begin{abstract}
We compare (i) nonfactorizable charm-quark loops in exclusive FCNC B-decays and (ii) three-particle contributions to the amplitude of semileptonic B-decay. 
Both amplitudes are given in the heavy-quark limit, $m_b\to\infty$, by a convolution of a hard kernel 
and a three-particle wave function of the $B$-meson,  
$\langle 0|\bar q(x)G_{\mu\nu}(z) b(0)|B(p)\rangle$.
An essential difference between the two amplitudes is that the amplitude of semileptonic $B$-decay involves this 3-particle wave function in a {\it collinear light-cone configuration}, whereas the amplitude of nonfactorizable charm in FCNC $B$-decays involves 
this 3-particle wave function in a {\it double collinear light-cone configuration}.  
\end{abstract}

\par
\keywords{Weak decays of B-mesons, FCNC, Meson distribution amplitudes in QCD\\[5pt]}

\maketitle
\thispagestyle{fancy}

\section{Introduction}\label{intro}
The purpose of our analysis is to study the contributions of 3-particle quark-antiquark-gluon states to amplitudes of different kinds of $B$-meson weak decays. We are going to identify those configurations of the 3-particle wave function of the $B$ meson that provide the dominant contributions to the amplitudes of 
(i) semileptonic (SL) $B$-decay and 
(ii) the nonfactorizable charming loops (NFcc) in flavour-changing neutral current (FCNC) $B$-decay. 

The basic object of our study is the amplitude of transition between the $B$-meson and the vacuum induced by the $T$-product of two bilinear quark currents $j$ and $J$ \cite{mnk2018}: 
\begin{eqnarray}
  A(p|q,q')=\int dx \exp(iqx)\langle 0|T\left\{j(x),J(0)\right\}|B(p)\rangle. 
\end{eqnarray}
The quark operators here are understood as Heisenberg operators of the Standard Model (i.e. the corresponding $S$-matrix contains both strong and weak interactions). We expand the $T$-product to the lowest necessary order in $G_F$ and the QCD coupling $\alpha_s$ when the 3-particle quark-antiquark-gluon contributions emerge. 

\subsection{Amplitude of SL \emph{B}-decay}

\begin{figure}[h]
\begin{center}
\includegraphics[height=5cm]{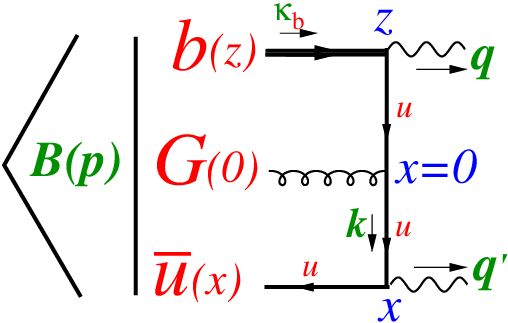}
\caption{\label{Fig:1}A diagram describing 3-particle contribution to the amplitude of semileptonic weak decay.}
\end{center}
\end{figure}

A semileptonic amplitude corresponds to the situation when one of the currents contains the $b$-quark field which may directly annihilate the $b$-quark in the $B$-meson: 
\begin{eqnarray}
  A_{\rm SL}(p|q,q')= \int dx \exp(i q' x)\langle 0|T\left\{\bar u(x)\gamma_\nu u(x),\bar u(0)\gamma_\mu b(0),
  \right\}|B_u(p)\rangle. 
\end{eqnarray}
The main contribution to the SL form factor is described by a well-known 2-particle quark-antiquark wave function of the $B$-meson. This contribution is beyond the scope of our interest. The 3-particle contribution of our interest \cite{offen2007,wang2022a,japan2001} comes into the game, when an additional soft gluon is emitted from the quark line (Fig.~\ref{Fig:1}):\footnote{The ($-i0$) in the propagators is always implied but is not written explicitly in most of the cases.}
\begin{eqnarray}
A_{\rm SL}(p|q,q')=\int \frac{dx dz}{(2\pi)^8} 
\frac{d\kappa_b\,e^{i \kappa_b z}}{m_u^2-(q-\kappa_b)^2-i0}\;\;
\frac{dk\, e^{-i k x+i q'x}}{m_u^2-k^2-i0}
\,\langle 0|\bar u(x)G(0)b(z)|B_s(p)\rangle. 
\end{eqnarray}
A specific feature of a diagram describing SL decay is that the heavy $b$-quark field hits the end-point of the propagator line (between the vertices emitting $q$ and $q'$) along which fast light degrees of freedom propagate \cite{m2022,m2023} 

\subsection{Amplitude involving charming loops in FCNC $B$-decay}
\begin{figure}[t]
\begin{center}
\includegraphics[height=5cm]{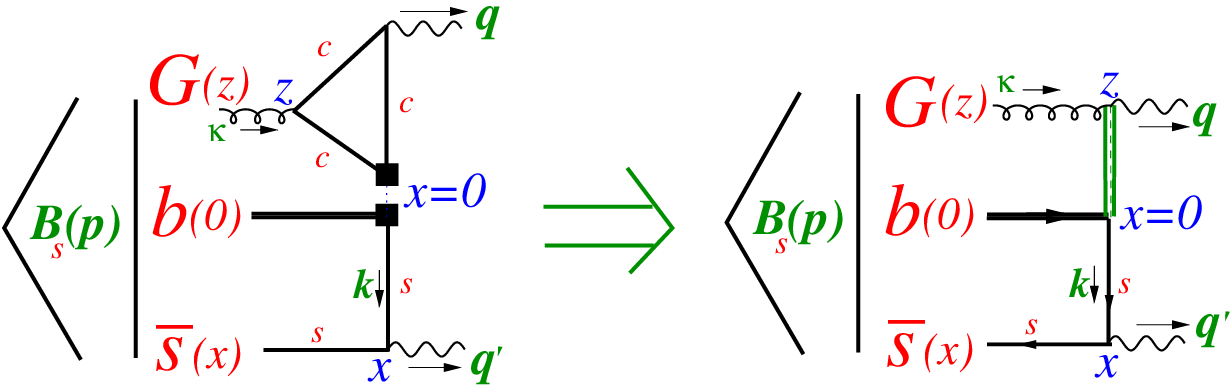}
\caption{\label{Fig:2}Reduction of the nonfactorizable charming loop to the diagram of generic weak form factor topology.}
\end{center}
\end{figure}

Charm-quark loops in FCNC amplitude correspond to the case when the currents do not contain the $b$-quark field to annihilate the $b$-quark in the $B$-meson: 
\begin{eqnarray}
  A_{\rm FCNC}(p|q,q')&=&\int dy \exp(i q y)\langle 0|T\left\{\bar c(y)\gamma_\mu c(y), \bar s(0)\gamma_\nu s(0)\right\}
  |B_s(p)\rangle
\end{eqnarray}
To obtain a nonzero result, one needs the insertion of a weak Hamiltonian describing the $b\to c\bar c s$ transition \cite{mk2018,m2019,Neubert,voloshin,ligeti,buchalla,khod1997,zwicky1,zwicky2,hidr,gubernari2020}.  So, the amplitude of our interest appears at order $G_F$ and contains $V_{cb}V^*_{cs}$. Hereafter, we omit these factors. The interesting 3-particle contribution is the one describing NFcc in FCNC $B$-decay (Fig.~\ref{Fig:2}) is defined as follows: 
\begin{eqnarray}
  A_{\rm NFcc}(p|q,q')=\int \frac{dx dz}{(2\pi)^4} d\kappa\,e^{i \kappa z}
  \Gamma_{cc}(\kappa, q)
\frac{dk\, e^{-i k x+i q'x}}{m_s^2-k^2}
\,\langle 0|\bar s(x)G(z) b(0)|B_s(p)\rangle, 
\end{eqnarray}
where $\Gamma_{cc}$ describes the charm-quark loop: 
\begin{eqnarray}
\label{tFeyn}
\Gamma_{cc}(\kappa, q)=\frac{1}{8\pi^2}
\int\limits_{0}^{1}du \int\limits_{0}^{1}dv 
\frac{\theta(u+v<1)}{m_c^2-uv (\kappa-q)^2 -u(1-u-v)\kappa^2-v(1-v-u)q^2}. 
\end{eqnarray}

The denominator of $\Gamma_{cc}$ is a quadratic function of the momenta, so $A_{\rm NFcc}$ is an example of the amplitude of the generic form factor topology, with the light-quark propagator replaced by an ``effective'' propagator
$\Gamma_{cc}(\kappa, q)$. 

The essential difference between the SL and NFcc amplitudes is the following: 
In the SL $B$-decay amplitude, the $b$-quark hits the end-point of the line along which fast light degrees of freedom propagate, while in the amplitude of NFcc the $b$-quark hits the middle point of this line \cite{mk2018,m2019,m2022,m2023}. 

We now study the consequences of this difference. For the sake of clarity, we can discuss the case of scalar "quarks" and "gluons", and we only need to distinguish between heavy ($\phi_b$) and light ($\phi$) fields. 

\section{3-particle contribution to an $\mbox{NFcc}$-type amplitude} 

\begin{figure}[t]
  \begin{center}
    \includegraphics[height=5cm]{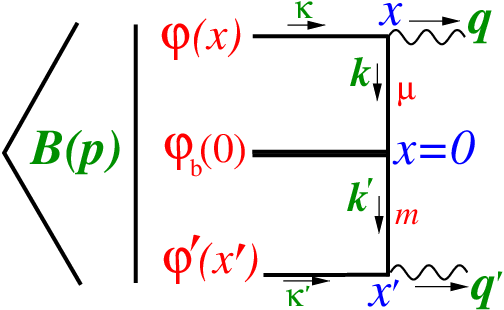}
    \caption{\label{Fig:3}Feynman diagram describing a simple  amplitude of a generic weak form factor topology, when the heavy constituent field hits the intermediate point of the light-field propagator line.}
    \end{center}
\end{figure}

The amplitude of our interest is shown in Fig.~\ref{Fig:3}. In \cite{m2025} amplitudes of this type are referred to as amplitudes of the generic form factor topology. 

In this case $\phi_b(0)$ is heavy and $\phi$ and $\phi'$ are light. 
The analytic expression for this amplitude has the form: 
\begin{eqnarray}
  \label{A0}
  A(p|q,q')=\int \frac{dx dx'}{(2\pi)^8}
  \frac{dk}{\mu^2-k^2}
  \frac{dk'}{m^2-k'^2}e^{i q x+i k x +i q' x'-i k' x'}
  \langle 0|\phi(x)\phi_b(0)\phi'(x')|B(p)\rangle. 
\end{eqnarray}

Let us derive the leading-order behavior of this amplitude for large $m_b$.
It is convenient to use light-cone coordinates: For any 4-vectors  
$a_\mu=(a_+,a_-,a_\perp)$, $b_\mu=(b_+,b_-,b_\perp)$, $ab=a_+b_-+a_-b_+-a_\perp b_\perp$, $a^2=2a_+a_--a_\perp^2$. 
We shall consider the case $q^2=q'^2=0$ and work in the rest frame of the $B$-meson. In this case we can take $q$ along $(+)$-axis, and $q'$ along the $(-)$-axis:  
\begin{eqnarray}
\label{refframe0}
p=(q_+,q'_-,0),\quad q=(q_+,0,0),\quad q'=(0,q_-,0),\quad 
q_+\propto M_B, \quad q'_-\propto M_B.     
\end{eqnarray}

\subsection{The $x$-vertex}
We introduce $\kappa=k+q$, the momentum carried by the constituent field $\phi(x)$: 
\begin{eqnarray}
 \int \frac{dx d\kappa}{\mu^2-(\kappa-q)^2}e^{i \kappa x}
 \langle 0|\phi(x)...|B(p)\rangle.
\end{eqnarray}
Due to properties of the $B$-meson wave function, the vector $\kappa$ is soft,
$\kappa_\mu\sim O(\Lambda_{\rm QCD})$, 
reflecting the fact that the heavy $b$ quark carries almost the full momentum of the heavy $B$-meson, whereas the light degrees of freedom carry its small fraction $O(\Lambda_{\rm QCD}/M_B)$.

The component $q_+\sim M_B$ is large and the propagator is highly virtual,
\begin{eqnarray}
\mu^2-2\kappa_-(\kappa_+-q_+)+\kappa_\perp^2\sim \Lambda_{\rm QCD}M_B.
\end{eqnarray}
Let us try to expand the field operator $\phi(x)$ near $x=0$. The expansion in powers of $x_-$ and $x_\perp$ leads
to a well-behaved Taylor series for the amplitude because
\begin{eqnarray}
  x_{-}e^{i\kappa_{+}x_{-}}
  \frac{1}{\mu^2-2\kappa_-(\kappa_+-q_+)+\kappa_\perp^2}
\to e^{i \kappa_+ x_-}\frac{\kappa_-}{(\mu^2-2\kappa_-(\kappa_+-q_+)+\kappa_\perp^2)^2}.
\end{eqnarray}
Since $\kappa_-=O(\Lambda_{\rm QCD})$ and the virtuality of the propagator is $O(\Lambda_{\rm QCD}M_B)$, 
any term $(x_-)^n$ is suppressed by a factor $(1/M_B)^n$ compared to the term $(x_-)^0$.
The same property holds for $(x_\perp)^n$.

However, for powers of the variable $x_+$ the situation is different: 
\begin{eqnarray}
  x_+ e^{i \kappa_- x_+}\frac{1}{\mu^2-2\kappa_-(\kappa_+-q_+)+\kappa_\perp^2}
  \to e^{i \kappa_- x_+}\frac{q_+}{(\mu^2-2\kappa_-(\kappa_+-q_+)+\kappa_\perp^2)^2}.
\end{eqnarray}
$q_+\sim M_B$, all powers of $x_+^n$ have the same order of magnitude $O(1)$: 
the Taylor expansion of $\phi(x_+)$ near $x_+=0$ does not lead to hierarchy 
in the corresponding expansion of the $B$-decay amplitude. 
We need to keep the full $x_+$-dependence of the operator $\phi(x_+)$ on the light cone ($x^2=0$).

So, the leading term of the expansion of the $B$-decay amplitude related to the $x$-vertex
corresponds to the expansion near $x_-=0,x_\perp=0$ and has the form 
\begin{eqnarray}
  \int \frac{dx_{+} dx_{-}dx_{\perp}}{(2\pi)^4} 
  \frac{d\kappa_+ d\kappa_- d\kappa_\perp}{\mu^2-2(\kappa_+-q_+)k_-+k_\perp^2}e^{i \kappa_+x_-+i \kappa_{-}x_+-ik_\perp x_\perp}
  \langle 0|...\phi(x_+)...|B(p)\rangle,
\end{eqnarray}
The integrals over $x_-$ and $x_\perp$ can be taken and lead to $\delta(\kappa_\perp)\delta(\kappa_+)$.
Integrating these $\delta$-functions, we obtain: 
\begin{eqnarray}
  \int \frac{dx_+ d\kappa_-}{2\pi}\frac{e^{i \kappa_-x_+}}{2 q_+ \kappa_--i0}
  \langle 0|...\phi(x_+)...|B(p)\rangle.
\end{eqnarray}
\subsection{The $x'$-vertex}
Now, $q'_-\sim M_B$ is the only nonzero component of the vector $q'$. The propagator has the form
\begin{eqnarray}
m^2-2\kappa'_-(\kappa'_--q'_-)+{\kappa'_\perp}^2\sim \Lambda_{\rm QCD}M_B.
\end{eqnarray}
Obviously, we can perform the Taylor expansion
of $\phi'(x')$ near $x'_+=0$ and $x'_\perp=0$ but we have to keep its full dependence on the variable $x_-$.
Taking into account this property and denoting $\tau'=x'_-$,
the dominant contribution of the $x'$-vertex reads
\begin{eqnarray}
  \int \frac{dx'_- d\kappa'_+}{2\pi}\frac{e^{i \kappa'_+x'_-}}{2 q'_- \kappa'_+ -i0}
  \langle 0|...\phi'(x'_-)...|B(p)\rangle.
  \nonumber
\end{eqnarray}

\subsection{The amplitude of Figure~\ref{Fig:3}}
Using the leading contributions of the $x$- and $x'$-vertices, we obtain 
\begin{eqnarray}
  \label{Apqfinal}
  A(p|q,q')=
  \int \frac{dx_+ d\kappa_-}{2\pi}
  \frac{e^{i \kappa_-x_+}}{2 q_+ \kappa_--i0}
  \int \frac{dx'_- d\kappa'_+}{2\pi}
  \frac{e^{i \kappa'_+x'_-}}{2 q'_- \kappa'_+-i0}
  \langle 0|\phi(x_+)\phi_b(0)\phi'(x'_-)|B(p)\rangle.
\end{eqnarray}
This is a particular case of {\it factorization theorem} [which has been proven for multiparticle contributions in \cite{m2025}; the name "double collinear" becomes fully clear within the general multiparticle case]: The dominant contribution to an amplitude of the generic weak form factor topology ($A_{\rm NFcc}$ represents such an amplitude) is given by the convolution of the hard kernel composed of propagators of light degrees of freedom and the 3-particle wave function in the ``double collinear'' kinematical configuration: the upper (above $b$-quark line) and the lower (below $b$-quark line) parts of the diagram are aligned along different LC directions.

\section{3-particle contributions to amplitudes of SL $B$-decays}
\begin{figure}[b]
\begin{center}
\includegraphics[height=5cm]{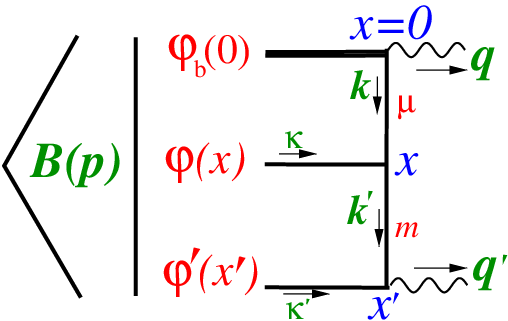} 
\caption{\label{Fig:4} Diagram for 3-particle contribution to weak SL form factor of $B$-decay.}
\end{center}
\end{figure}

The three-particle contribution to the amplitude of SL topology of Fig.~\ref{Fig:4} reads: 
\begin{eqnarray}
A_{\rm SL}(p\,|q,q')=\int \frac{dx'\, dx}{(2\pi)^8}
\frac{d\kappa\, e^{i \kappa x}}{\mu^2-(q'-\kappa'-\kappa)^2}\;\;
\frac{d\kappa'\,e^{i\kappa'x'}}{m^2-(q'-\kappa')^2}
\langle 0|\varphi_b(0)\varphi(x)\varphi'(x')|B(p)\rangle.
\end{eqnarray}

We introduce the Feynman parameter $v$ and redefine the variables as follows
\begin{eqnarray}
x=v\,x' +z, \quad 
\tilde\kappa'=\kappa'+ v\kappa.
\end{eqnarray}
Here, $z$ measures the deviation of $x$ from the straight line joining the end points $0$ and $x'$. In the new variables 
\begin{eqnarray}
\label{ASL3}
A_{\rm SL}(p|q,q')&=&\int\limits_0^1 \frac{dv}{(2\pi)^8}\int 
\frac{ dz\,dx'\,d \kappa \,d\tilde\kappa'\,  
e^{i \tilde\kappa' x'+i\kappa z}}
{\left[m^2(1-v)+\mu^2 v 
-(q'-\tilde\kappa')^2
-v(1-v){\kappa^2}\right]^2}\nonumber\\
&&\hspace{2cm}\times\langle 0|\varphi_b(0)\varphi(x' v+{z})\varphi'(x')|B(p)\rangle.
\end{eqnarray}
Since $\kappa_\mu,\,\tilde\kappa'_\mu=O(\Lambda_{\rm QCD})$
are soft compared to 
$(q'-\tilde\kappa')^2=O(m_b \Lambda_{\rm QCD})$,
we can set $\kappa\to 0$ in the denominator, so that the $\kappa$-integration leads to $\delta(z)$.
So we conclude that the dominant contribution to $A_{\rm SL}(p|q,q')$ comes from the collinear kinematical configuration of 3-particle wave function, when the points $x=0$, $x$, and $x'$ lie on a straight line, with $x$ between the end points $x=0$ and $x'$: $x_\mu=v x'_\mu$, $0<v<1$.

In the reference frame (\ref{refframe0}), the amplitude (\ref{ASL3}) takes the following form for large $m_b$:  
\begin{eqnarray}
\label{ASLfinal}
A_{\rm SL}(p|q,q')&=&\int\limits_0^1 dv 
\int \frac{dx'_- d\tilde k'_+}{2\pi}
\frac{e^{ix'_-\tilde k'_+}}
{(2\tilde k'_+q'_--i0)^2}
\langle 0|\varphi_b(0)
\varphi(v\,x'_- )
\varphi'(x'_-)|B(p)\rangle.
\end{eqnarray}
\section{Multi-particle contribution to amplitudes of \emph{B}-decays}
\begin{figure}[h!]
\begin{center}
\includegraphics[height=5cm]{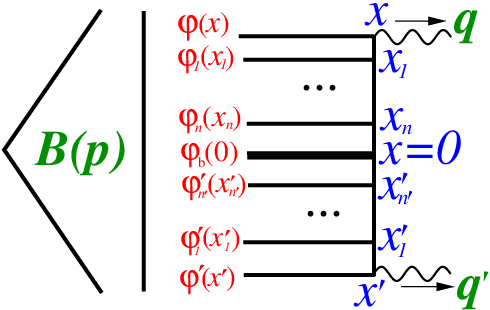}
\caption{\label{Fig:5} A diagram of the generic weak form factor topology \cite{m2025}.}
\end{center}
\end{figure}

The results of the previous sections may be generalized 
to the case of multi-particle contributions to $B$-decay amplitudes of the generic form factor topology 
of Fig.~\ref{Fig:5}, see \cite{m2025} for details. The corresponding $B$-decay amplitude involves the multi-particle wave function of the $B$-meson (the field $\phi_b$ is heavy, all other fields are light): 
\begin{eqnarray}
  \langle 0|\phi(x)\phi_1(x_1)\ldots \phi_n(x_n) \phi_b(0)
  \phi'_{n'}(x'_{n'})\ldots\phi_1'(x_1')\phi'(\tau')|B(p)\rangle.
\end{eqnarray}

In the reference frame where $q$ lies along the $(+)$-direction and $q'$ lies along the $(-)$-direction,
\begin{eqnarray}
  q=(q_+,0,0),\quad 
  q'=(0,q'_-,0),\qquad q_+=q'_-\propto M_B, 
\end{eqnarray}
the dominant contribution to the amplitude of $B$ decay comes from the following configuration:
\begin{eqnarray}
& x=(\tau,0,0),    \quad & x_1=(\tau u_1,0,0),\quad \,\,\ldots,\quad     x_n=(\tau u_n,0,0),   \qquad 0<u_n<\ldots<u_1<1, \nonumber\\
&  x'=(0,\tau',0), \quad & x'_1=(0,\tau' u'_1,0),\quad \ldots,\quad x'_{n'}=(0,\tau' u'_{n'},0), \,\quad 0<u'_{n'}<\ldots<u'_1<1.
\end{eqnarray}
The coordinates $x,x_1,\ldots,x_n$ are ordered and lie on the $(+)$-axis of the LC (i.e., along the large component $q_+$) 
and the coordinates $x',x'_1,\ldots,x'_{n'}$ are ordered and lie on the $(-)$-axis of the LC (i.e., along the large component $q'_-$). That is why we refer to this configuration as the double collinear LC configuration. 

\section{Summary}
We studied a $B$-decay amplitude of a generic form factor topology (an amplitude given by diagrams in which the heavy
field hits an intermediate point of the line along which fast light degrees of freedom propagate) and obtained the following results: 

\vspace{.2cm}
\noindent 
$\bullet$ The 3-particle contribution to the amplitudes describing nonfactorizable charming loops in FCNC $B$ decays, $A_{\rm NFcc}$,
may be calculated in the $B$-meson rest frame specified by the conditions:  
\begin{eqnarray}
\label{refframe}
q_\mu \propto a_\mu, \quad q'_\mu \propto a'_\mu,
\quad a^2=0,\quad  a'^2=0, \quad aa'=2,  
\end{eqnarray}
as the convolution of (i) the hard kernel (hard propagators) and (ii) 
the three-particle quark-antiquark-gluon wave function of the $B$-meson in a {\it double collinear LC configuration}
\begin{eqnarray}
\langle 0|\bar q(\tau a_\mu)b(0)G_{\mu\nu}(\tau' a'_\mu)|B(p)\rangle.
\end{eqnarray}
Eq.~(\ref{Apqfinal}) gives the corresponding convolution formula; corrections to this formula are suppressed by powers of $m_b$. 
The application of double-collinear 3-particle wave functions to FCNC $B$-decays in QCD were presented in \cite{wang2022,wang2023,bbm2023,bbm2024}. 

\vspace{.2cm}
\noindent 
$\bullet$ The 3-particle contributions to the amplitude of SL $B$-decays, $A_{\rm SL}$, may be calculated in the reference frame 
(\ref{refframe}) as the convolution of (i) the hard kernel and (ii) the three-particle quark-antiquark-gluon wave function of the $B$-meson in a {\it collinear LC configuration}:
\begin{eqnarray}
\langle 0|\bar q(\tau a'_\mu)b(0)G_{\mu\nu}(v \tau a'_\mu)|B(p)\rangle,\quad 0 < v < 1. 
\end{eqnarray}
Corrections to the convolution formula are suppressed by powers of $m_b$. 

We would like to emphasize that the use of the collinear 3-particle wave function for the description of nonfactorizable charming loops in FCNC $B$-decays \cite{hidr,gubernari2020} does not seem to us theoretically justified.

\begin{acknowledgments}
I am grateful to Ilya Belov and Alexander Berezhnoy for a fruitful collaboration on this topic. 
\end{acknowledgments}

\section*{FUNDING}
This study was conducted under the state assignment of Lomonosov Moscow State University. 

\section*{CONFLICT OF INTEREST}
The author of this work declares that he has no conflicts of interest.




\begin{thebibliography}{100}
\bibitem{mnk2018}
A.~Kozachuk, D.~Melikhov, and N.~Nikitin,
{\it Rare FCNC radiative leptonic $B_{s,d}\to \gamma l^+l^-$ decays in the standard model},
Phys. Rev. D \textbf{97}, 053007 (2018).
\bibitem{offen2007}
A.~Khodjamirian, T.~Mannel, and N.~Offen,
{\it Form-factors from light-cone sum rules with B-meson distribution amplitudes},
Phys.~Rev.~{\bf D75}, 054013 (2007).

\bibitem{wang2022a}
B.~Y.~Cui, Y.~K.~Huang, Y.~L.~Shen, C.~Wang and Y.-M.~Wang,
{\it Precision calculations of $B_{d,s}\to (\pi, K)$ decay form factors in soft-collinear effective theory}, 
JHEP \textbf{03}, 140 (2023). 

\bibitem{japan2001}
H.~Kawamura, J.~Kodaira, C.-F.~Qiao, and K.~Tanaka, 
{\it B-meson light cone distribution amplitudes in the heavy quark limit}, 
Phys.~Lett.~{\bf B523}, 111 (2001), Erratum: Phys.~Lett. {\bf B536}, 344 (2002).

\bibitem{m2022}
D.~Melikhov,
{\it Nonfactorizable charming loops in FCNC B decay versus B-decay semileptonic form factors}, Phys.~Rev.~{\bf D106}, 054022 (2022).

\bibitem{m2023}
D.~Melikhov,
{\it Three-particle distribution in the B meson and charm-quark loops in FCNC B decays},
Phys.~Rev.~{\bf D108}, 034007 (2023).

\bibitem{mk2018}
A.~Kozachuk and D.~Melikhov, 
{\it Revisiting nonfactorizable charm-loop effects in exclusive FCNC $B$ decays},
Phys.~Lett.~{\bf B786}, 378 (2018).
\bibitem{m2019}
D.~Melikhov,
{\it Charming loops in exclusive rare FCNC $B$-decays}, EPJ Web Conf. {\bf 222}, 01007 (2019).
\bibitem{Neubert}
M.~Beneke, G.~Buchalla, M.~Neubert, and C.~T.~Sachrajda,
{\it Penguins with Charm and Quark-Hadron Duality}, 
Eur.~Phys.~J. {\bf C61}, 439 (2009).

\bibitem{voloshin}
M.~B.~Voloshin, {\it Large $O(m_c^{-2})$ nonperturbative correction
to the inclusive rate of the decay $B\to X_s\gamma$},  
Phys.~Lett. {\bf B397}, 275 (1997). 

\bibitem{ligeti}
Z.~Ligeti, L.~Randall, and M.~B.~Wise, 
{\it Comment on nonperturbative effects in $\bar B\to X_s\gamma$},
Phys.~Lett. {\bf B402}, 178 (1997). 

\bibitem{buchalla}
G.~Buchalla, G.~Isidori, S.~J.~Rey, 
{\it Corrections of order $\Lambda_{\rm QCD}^2/m_c^2$ to inclusive rare $B$ decays}, 
Nucl.~Phys. {\bf B511}, 594 (1998). 

\bibitem{khod1997}
A.~Khodjamirian, R.~Ruckl, G.~Stoll, and D.~Wyler, 
{\it QCD estimate of the long distance effect in $B\to K^*\gamma$}, 
Phys.~Lett. {\bf B402}, 167 (1997). 

\bibitem{zwicky1}
P.~Ball and R.~Zwicky, 	
{\it Time-dependent CP Asymmetry in $B\to K^*\gamma$ as a (Quasi) Null Test of the Standard Model}, 
Phys.~Lett. {\bf B642}, 478 (2006). 

\bibitem{zwicky2}
P.~Ball, G.~W.~Jones, and R.~Zwicky, 	
{\it $B\to V\gamma$ beyond QCD factorisation}, 
Phys.~Rev.~{\bf D75}, 054004 (2007). 

\bibitem{hidr}
A.~Khodjamirian, T.~Mannel, A.~Pivovarov, and Y.-M.~Wang, 
{\it Charm-loop effect in $B\to K^{(*)} l^+l^-$ and $B\to K^*\gamma$}, 
JHEP {\bf 09}, 089 (2010).

\bibitem{gubernari2020}
N.~Gubernari, D.~van Dyk, J.~Virto, 
{\it Non-local matrix elements in $B_{(s)}\to \{K^{(*)},\phi\}\ell^+\ell^-$},
JHEP {\bf 02}, 088 (2021).

\bibitem{m2025}
D.~Melikhov, 
{\it Factorization of multiparticle contributions to amplitudes of B-meson weak decays}, e-Print: 2507.23441 [hep-ph]
\bibitem{wang2022}
Q.~Qin, Y.~L.~Shen, C.~Wang and Y.~M.~Wang,
{\it Deciphering the long-distance penguin contribution to $\bar B_{d, s} \to \gamma \gamma$ decays},
Phys.~Rev.~Lett. {\bf 131}  091902 (2023).
\bibitem{wang2023}
Y.-K. Huang, Y.~Ji, Y.-L.~Shen, C.~Wang, Y.-M.~Wang, X.-C.~Zhao, 
{\it Renormalization-Group Evolution for the Bottom-Meson Soft Function},
Phys.~Rev.~Lett. {\bf  133}, 171901 (2024).
\bibitem{bbm2023}
I.~Belov, A.~Berezhnoy, and D.~Melikhov,
{\it Charming-loop contributions in $B_s\to \gamma\gamma$ decay},
Phys.~Rev.~{\bf D108}, 094022 (2023).
\bibitem{bbm2024}
I.~Belov, A.~Berezhnoy, and D.~Melikhov,
{\it Nonfactorizable charming-loop contribution to FCNC $B_s\to \gamma ll$ decay}, 
Phys.~Rev.~{\bf D109}, 114012 (2024). 

\end{thebibliography}
\end{document}